\documentclass[11pt]{article}
\usepackage{latexsym}
\begin{document}
\newcommand{\roughly}[1]%
       
%{\mathrel{\raise.4ex\hbox{$#1$\kern-.75em\lower1ex\hbox{$\sim$}}}}
\newcommand{\PSbox}[3]{\mbox{\rule{0in}{#3}
\includegraphics{#1}\hspace{#2}}}
\newcommand\lsim{\roughly{<}}
\newcommand\gsim{\roughly{>}}
\newcommand\CL{{\cal L}}
\newcommand\CO{{\cal O}}
\newcommand\half{\frac{1}{2}}
\newcommand\beq{\begin{eqnarray}}
\newcommand\eeq{\end{eqnarray}}
\newcommand\eqn[1]{\label{eq:#1}}
\newcommand\intg{\int\,\sqrt{-g}\,}
\newcommand\eq[1]{eq. (\ref{eq:#1})}
\newcommand\meN[1]{\langle N \vert #1 \vert N \rangle}
\newcommand\meNi[1]{\langle N_i \vert #1 \vert N_i \rangle}
\newcommand\mep[1]{\langle p \vert #1 \vert p \rangle}
\newcommand\men[1]{\langle n \vert #1 \vert n \rangle}
\newcommand\mea[1]{\langle A \vert #1 \vert A \rangle}
\newcommand\bi{\begin{itemize}}
\newcommand\ei{\end{itemize}}
\newcommand\be{\begin{equation}}
\newcommand\ee{\end{equation}}
\newcommand\bea{\begin{eqnarray}}
\newcommand\eea{\end{eqnarray}}

\def\Dsl{\,\raise.15ex \hbox{/}\mkern-12.8mu D}
\newcommand\Tr{{\rm Tr\,}}
\thispagestyle{empty}
\begin{titlepage}
\begin{flushright}
CALT-68-2344\\
\end{flushright}
\vspace{1.0cm}
\begin{center}
{\LARGE \bf  Forecasting Portfolio Risk}\\ 
\bigskip
{\LARGE \bf in Normal and Stressed Markets}\\
~\\
\bigskip\bigskip
{ Vineer Bhansali$^a$ and Mark B. Wise$^b$} \\
~\\
\noindent
{\it\ignorespaces
          (a) PIMCO, 840 Newport Center Drive, Suite 300\\
               Newport Beach, CA 92660 \\

          {\tt bhansali@pimco.com}\\
\bigskip   (b) California Institute of Technology, Pasadena CA 91125\\

{\tt   wise@theory.caltech.edu}
}\bigskip
\end{center}
\vspace{1cm}
\begin{abstract}
The instability of historical risk factor correlations renders their use in estimating portfolio risk extremely questionable.
In periods of market stress correlations of risk factors have a tendency to quickly go well beyond estimated values. For instance, in times of severe market stress, one would expect with certainty to see the correlation of yield levels and credit spreads to go to -1, even though historical estimates will miss this region of correlation. 
This event might lead to realized portfolio risk profile substantially different than what was initially estimated. The purpose of this paper is to explore the effects of correlations on fixed income portfolio risks. To achieve this,
we propose a methodology to estimate portfolio risks in both normal and stressed times using confidence weighted forecast correlations. 
\end{abstract}
\vfill
%\today
\end{titlepage}
%\maketitle

\section{Background}

Most calculations of portfolio risk require an estimate of the volatilities and correlations of the assets in the portfolio. The traditional methodology assumes that correlations and volatilities obtained from historical data are a fair estimate of future correlations and volatilities. Some sophisticated techniques, as summarized in Litterman and Winkelmann (1998) weight recent data more heavily, but still suffer from the basic problem of depending too heavily on history. Most standard techniques for estimating Value at Risk\footnote {See for example, Jorion (2001). For a recent discussion of some sources of bias in VAR estimates see, Xiongwei and Pearson (1999).}(VaR) assume that the distribution of returns of asset prices in the past will hold into the future, hence an estimate of the tail of the return distribution would yield a good estimate of the loss in value of the portfolio in times of stress. Obtaining implied correlations from traded option prices is possible in foreign exchange and interest rates to some degree, but far from satisfactory when multiple asset classes are involved (see, for example Bhansali (1997)). 

At some firms reliance on historical estimates of correlation and volatility are treated with skepticism, because of the simple fact that these historical estimates fail miserably in times of market stress, and even in normal times are at best quite inaccurate. Correlation and volatility forecasts that are arrived at as part of a qualitative process involving discussions and meetings are then considered more trustworthy. 

The ratio of expected excess return to expected volatility is the Sharpe ratio of a portfolio\footnote {See, for example,  Wilmott (2000).}. As an active bond manager, one cannot consistently obtain an estimate of the ex-ante Sharpe ratio by forecasting excess returns (the numerator) and using historical data for measuring risk (the denominator). What one would like to do is also forecast correlations and volatilities that are expected to be realized in different market scenarios (with levels of certainty for different correlations and volatilities based on our confidence in the forecasts) and then estimate the Sharpe ratio for the future.

If there are many different risk factors, making up a correlation matrix is not simple. An example will show the inherent problem. Assume that we have three risk factors, level of yields, slope of the yield curve and spreads of non-government sector. Then, as soon as the correlation between level and slope, and the correlation between level and spread, are specified the correlation between slope and spread is automatically restricted to be in a region whose range depends on the
specified correlations.  This range is determined by the requirement that the correlation matrix be positive semi-definite. For a large set of factors, this problem is compounded, {\it i.e.} selecting a set of correlations by hand restricts the ranges of all the other correlations in a very complicated way. Typically a forecasted
correlation matrix obtained by some qualitative process will not be mathematically consistent and a methodology must be introduced to obtain a consistent correlation matrix from the forecasted one.

A group of portfolio managers are likely to have a high degree of confidence in forecasting specific elements of the correlation matrix, {\it e.g.} the correlation between level and slope, and slightly lower degree of confidence in forecasting other correlations, {\it e.g.} the correlation between level and spreads. The methodology introduced
by Rebonato and J\"{a}ckel (2000) provides a convenient way to obtain a consistent correlation matrix that is close to the original forecast. Elements that we have a lot of confidence about are forced to be close to the desired values using a systematic weighting procedure, and more freedom is allowed in the selection of the elements about which we are not so certain. The confidence weighting is particularly important for large correlation matrices, where forecasting all of the elements is not practical.  Taking most of the elements from historical data and forecasting a few of the key elements, the weighting procedure can be arranged to get from this hybrid historical/forecasted
matrix  a consistent correlation matrix with elements that are very close to the forecasted ones (however, some of the elements taken from historical data may change significantly). 

The purpose of this paper is to illustrate with some simple examples the importance (particularly in times of market stress) 
of using forecast correlation matrices and also
to elaborate on the method of Rebonato and J\"{a}ckel (2000) for obtaining consistent confidence weighted correlation matrices that are as close as possible to the forecasted one. One small technical improvement on the work of Rebonato and J\"{a}ckel
is the use of a parameterization for the most general $N \times N$ correlation matrix that involves only $N(N-1)/2$ angles. 

In the next section we illustrate, with two examples, how forecast correlation matrices change between normal and stressed economic environments. The problem of mathematical consistency of forecast correlation matrices is also discussed. Section 3 elaborates on the method of Rebonato and J\"{a}ckel for obtaining a consistent confidence weighted correlation matrices from a forecasted matrix. In section 4 we illustrate the importance of forecasting correlation matrices by examining how correlations effect the total level duration of portfolios in normal and stressed economic environments. Finally some concluding remarks are given in section 5.

\section{Factors and Correlations}

Let us begin by identifying some of the main sources of risk for typical fixed income portfolios. This list is by no means exhaustive, and used solely for illustration. 

\bi

\item Duration: Risk due to the change in yield level factor (Level). 
\item 2-10 Duration: Risk due to the change in slope factor between the 2 and 10 year points (${\rm Slope_{2-10}}$).\footnote{In many risk-management systems the risk of yield curve reshapings is measured by ``key-rate" durations in place of the $2-10$ and $10-30$ durations.}
\item 10-30 Duration: Risk due to the change in slope factor between 10 and 30 year points (${\rm Slope_{10-30}}$).
\item Mortgage Spread Duration: Risk due to the change in the mortgage spread factor (Mortgage) measured against the benchmark treasury curve.
\item Corporate Spread Duration: Risk due to the change in the corporate spread factor (Corporate) measured against the benchmark treasury curve.

\ei

This list can be expanded to include risk factors such as currency, implied tax rates (for municipals), convertibles, implied inflation rates (for TIPS), EMBI spread (for emerging market bonds) etc.   For purpose of illustrating our approach in complete detail, we will work only with the five factors listed above.

One possible expectation for the  signs of the elements of the correlation matrix is
\be
\bordermatrix{&{\rm Level}&{\rm Slope_{2-10}}&{\rm Slope_{10-30}}&{\rm Mortgage}&{\rm Corporate}\cr
	{\rm Level}&1&-&-&-&-\cr
	{\rm Slope_{2-10}}&-&1&+&+&+\cr
	{\rm Slope_{10-30}}&-&+&1&+&+ \cr
	{\rm Mortgage}&-&+&+&1&+ \cr
	{\rm Corporate}&-&+&+&+&1 }.
\ee

Thus, for example as the yields fall we expect corporate spreads and mortgage spreads to rise ({\it i.e.} a negative correlation of corporate and mortgage spreads with yield level). Similarly in an easing cycle as yields fall, we would expect the curve to steepen (again a negative correlation of yield level with the slope parameters). Of course, these correlations can change and do change depending on the environment. For example it is entirely possible that as mortgage spreads narrow corporate spreads narrow in most environments but widen in some. A good example of corporate spreads and mortgage spreads moving in opposite directions is provided by the differential performance of premium coupon Fannie Mae mortgages pass-throughs and interest rate swaps this year. Fannie Mae pass-through OASes (option adjusted spreads) are a good measure of the relative value in the mortgage market. As the level of rates has fallen this year, the increase in refinancing activity has adversely affected the spreads on mortgages that are more likely to prepay, i.e. those with high coupons. On the other hand, the Fed's active reduction of rates has resulted in an over-all reduction in the level of swap-spreads, which is a good measure of corporate spreads for the financial sector.

To estimate portfolio risk, we have to also assign magnitudes to these correlations. So a reasonable guess in today's environment, assuming things remain reasonably stable over the horizon would be a correlation matrix:

\be
\label{normalmat}
\bordermatrix{&{\rm Level}&{\rm Slope_{2-10}}&{\rm Slope_{10-30}}&{\rm Mortgage}&{\rm Corporate}\cr
	{\rm Level}&1&-0.50&-0.30&-0.25&-0.70\cr
	{\rm Slope_{2-10}}&-0.50&1&0.90&0.30&0.70\cr
	{\rm Slope_{10-30}}&-0.30&0.90&1&0.25&0.20 \cr
	{\rm Mortgage}&-0.25&0.30&0.25&1&0.75 \cr
	{\rm Corporate}&-0.70&0.70&0.20&0.75&1 }.
\ee

What might happen to these correlations in severe market distress?  History shows that in periods of distress the Fed eases aggressively, so short maturity yields fall, the yield curve steepens aggressively and spreads widen out\footnote{The transmission mechanism for the drastic change in the correlations is probably through the reduction in ``liquidity" . Note that it is assumed in our paper that in periods of stress the underlying relative volatility of the risk factors themselves does not change substantially, {\it i.e.} that the correlation changes carry most of the transition of the covariance matrix.}(mortgages and corporates cheapen).  In such times the correlation matrix would be expected to look like:

\be
\label{stressmatnpd}
\bordermatrix{&{\rm Level}&{\rm Slope_{2-10}}&{\rm Slope_{10-30}}&{\rm Mortgage}&{\rm Corporate}\cr
	{\rm Level}			&	1	&	-0.90	&	-0.55	&	-1	&	-1\cr
	{\rm Slope_{2-10}}	&-0.90	&1		& 0.98	& 0.90		& 0.90\cr
	{\rm Slope_{10-30}}	&-0.55	& 0.98	&1		& 0.55	& 0.55 \cr
	{\rm Mortgage}		&-1		& 0.90		& 0.55	&1		& 1 \cr
	{\rm Corporate}		&-1		& 0.90		& 0.55	& 1	&1 } ,
\ee
{\it i.e.} all the correlations have increased in magnitude and some have been driven to their extreme values (see, for
example, Boyer, Gibson and Loretan (1997)).

Naively both these correlation matrices look reasonable. They are inconsistent, however, as these correlations are mathematically impossible. There is no way the correlations that we guessed in equations (\ref{normalmat}) or (\ref{stressmatnpd}), can ever be observed in real data.  The technical statement of this fact is that the eigenvalues of these correlation matrices are not all non-negative. In economic terms, this translates to the fact that if such  correlation matrices were allowed, one would have to allow for negative variance, or imaginary volatility, which is impossible.\footnote{The eigenvalues of the matrix in equation (\ref{stressmatnpd}) are 4.36, 0.73, -0.10, 0, 0. Two of the eigenvalues are zero because some of the correlations are at their extreme vales. In the next section we discuss in geometric terms the origin of this inconsistency. The eigenvalues for the matrix in equation (\ref{normalmat}) are 2.99, 1.16, 0.74, 0.24,-0.13.}

The fact that a forecast correlation matrix is not mathematically consistent and has negative eigenvalues is not necessarily of crucial importance. For
example the total level durations calculated in section 4 for various sample portfolios do not change very much if the forecast correlation matrices above are used or the mathematically consistent ones introduced in the next section that are close to those in
equations (\ref{normalmat}) and (\ref{stressmatnpd}) are used. However, using a mathematically consistent correlation matrix is very important if the portfolio is optimized to minimize risk based on the correlation matrix.

\section{Consistent Correlation Matrices}

An $N \times N$ correlation matrix $\rho$ is positive definite and it therefore can be written as
\be
\label{defv}
\rho_{ij}~=~ \sum v_{ik}v_{jk},
\ee 
where the column index $k$, of the $N \times N$ matrix $v$, is summed from 1 to $N$. Since the
diagonal elements of the correlation matrix are equal to one,
each row of the matrix $v$ is a unit vector in an $N$ dimensional space with the
elements of the row being the components of that vector. Thus elements
of the correlation matrix have a simple geometric interpretation. They
are the cosines of the angles between these unit vectors.  When the $ij$ (off diagonal) element of the correlation matrix equals its extreme value ({\it i.e.} $\pm 1$) then the $i$'th and $j$'th vectors are either parallel or anti-parallel. If the $ij$ (off diagonal) element of the correlation matrix is zero then the $i$'th and $j$'th vectors are perpendicular.

Using this geometric point of view it is easy to understand why the correlation matrix in
equation (\ref{stressmatnpd}) is inconsistent.  
The ${\rm Slope_ {2-10}}-{\rm Slope_{10-30}}$ element of the
correlation matrix is very close to unity and so the ${\rm Slope_{2-10}}$ and ${\rm Slope_{10-30}}$ vectors
are almost parallel. However with them almost parallel it is impossible for the angles between them
and the ${\rm Level}$ vector, the Mortgage vector and the Corporate vector to be so different. 

A rotation\footnote{For rotations $O$ is an orthogonal matrix and satisfies $O^T O=1$.}of the components of all
the vectors, $v_{ik} \rightarrow O_{kp}v_{ip}$, does not change the correlation matrix. By making such
rotations it is possible to align the vectors so that $v_{1k}=0$ for $k=2,..., N$, $v_{2k}=0$ for $k=3,..., N$, {\it etc}.
Hence the unit vectors can be expressed in terms of $N(N-1)/2$ angles which specify their relative orientations. Explicitly, for the $N=5$ case
considered in the previous section, we take,
\bea
\label{angles}
v_{1k}&=&(1,0,0,0,0),  \nonumber\\
v_{2k}&=&(c_1,s_1,0,0,0), \nonumber \\
v_{3k}&=&(c_2,s_2 c_3,s_2 s_3,0,0), \\
v_{4k}&=&(c_4,s_4 c_5,s_4 s_5 c_6,s_4 s_5 s_6,0), \nonumber\\
v_{5k}&=&(c_7,s_7 c_8,s_7 s_8 c_9,s_7 s_8 s_9 c_{10},s_7 s_8 s_9 s_{10}), \nonumber 
\eea
where we have adopted the simplifying notation $c_i= {\rm cos}\theta_i$ and $s_i= {\rm sin}\theta_i$.
The most general $5\times 5$ consistent correlation matrix, $ \rho^{(g)} = v v^T$, is then a function of the ten angles $\theta_1,...,\theta_{10}$.  The parameterization above is a little different from that proposed in
Rebonato and J\"{a}ckel (2000). We have used the rotational invariance to reduce the number of parameters in
an $N\times N$ correlation matrix from $N(N-1)$ to $N(N-1)/2$.

When forecasting the correlation matrix, we usually have a higher degree of confidence in some of the entries. For example, we could be much more confident that we know the correlation between level and the ${\rm 2-10}$ slope or between level and corporate spreads, as compared to the other correlations. This could arise from the historically observed tendency of ``bull-steepeners", i.e. Fed easing in the short end is usually accompanied with a fall in yields across the yield curve and expectations of spread product outperformance. We encapsulate these prejudices using a (symmetric)
``confidence matrix" $C$; the larger the value of
the element $C_{ij}$ the more confidence we have in the $ij$ element of the forecasted correlation matrix $\rho^{(f)}$. Practically, the confidence matrix can be arrived at by either expressing a strong macroeconomic view, or by averaging the weighted forecasts of a number of portfolio managers. 
The forecasted correlation matrix $\rho^{(f)}$ may not be allowed because it has some negative eigenvalues. A consistent correlation matrix $ \rho^{(c)} $, appropriate for the level of confidence we have in the various elements of $\rho^{(f)}$, is constructed by choosing the angles it depends on to minimize the function \footnote { For some other approaches to the problem of obtaining consistent correlation matrices see Finger (1997) and Kupiec (1998).}(Rebonato and J\"{a}ckel (2000))

\be
\label{method}
f=\sum C_{ij}(\rho^{(c)}_{ij}-\rho^{(f)}_{ij})^2,
\ee
where the sum goes over all values of $i$ and $j$ from $1$ to $N$. Since $\rho_{ii}=1$
the diagonal elements of the confidence matrix are irrelevant and we adopt the convention that
the diagonal elements of $C$ are zero. So, for example,
if the ${\rm Level- Slope_{2-10}}$ element of the confidence matrix was 100 and its other off diagonal elements were unity we
would expect to find the difference between $\rho^{(c)}_{L,S_{2-10}}$ and $\rho^{(f)}_{L,S_{2-10}}$ to be ten times smaller in magnitude than the typical difference between elements of the consistent correlation matrix and the forecasted one.

 There are
some situations where this will not be the case. For example, suppose that the off diagonal 
elements of the confidence matrix
have either the value 1 or 100. If the elements of  $\rho^{(f)}$ corresponding to those of $C$ that have the value 100
are not part of a consistent correlation matrix for any value of the other elements of $\rho^{(f)}$ then, despite the
large values of these elements of $C$, some of the corresponding elements of $\rho^{(c)}$ will differ significantly
from those of $\rho^{(f)}$. In this case the enhanced confidence in these elements of $\rho^{(f)}$ was a priori misguided.

We could always have high confidence in any row of $\rho^{(f)}$ without
running into this problem. This is evident from the geometric interpretation of the correlation matrix. It
is always possible to specify any angle between $N-1$ unit vectors and some arbitrary axis. 
There is a simple diagrammatic way to know if a particular set of elements of $\rho^{(f)}$
can be chosen with high confidence (for any values of those elements). Draw $N$ circles labeled by the numbers $1,..., N$.
Then for each of the elements $\rho^{(f)}_{ij}$ , that is in the high confidence set,
draw a line that joins circle $i$ with circle $j$. If the resulting diagram has no closed
loops then this set of elements can always be chosen freely with high confidence. To prove this assertion suppose the confidence diagram contains the closed loop $[1,...,i]$, which denotes that circle 1 is joined by a line to circle 2, etc. The closed loop occurs because circle i is also joined by a line to circle 1. Now imagine that the elements $\rho_{12}^{(f)}=...=\rho_{i-1,i}^{(f)}=1$. Then we can no longer choose freely the
element $\rho_{i-1,i}^{(f)}$. It must also be unity since the vectors $1,...,i$ are all parallel. A diagram with no closed loops is composed of segments of the form $(1,...,i)$, which is similar to the
square bracket case above except that circle i is not joined to circle 1 by a line. It is always possible to
choose $i$ vectors with the angles between vectors 1 and 2, ... ,$i-1$ and $i$ fixed at any values.

Lets apply these ideas to the forecasted correlation matrices in equations (\ref{normalmat}) and (\ref{stressmatnpd}).
Suppose the confidence matrix $C$ was 
\be
\label{weightmat}
\bordermatrix{&{\rm Level}&{\rm Slope_{2-10}}&{\rm Slope_{10-30}}&{\rm Mortgage}&{\rm Corporate}\cr
	{\rm Level}			&	0	&	100	&1		&	100	&	100\cr
	{\rm Slope_{2-10}}	&100		&	0	&100		&1		&1\cr
	{\rm Slope_{10-30}}	&1		&100		&	0	&1		&1 \cr
	{\rm Mortgage}		&100		&1		&1		&	0	&1 \cr
	{\rm Corporate}		&100		&1		&1		&1	& 0 }.
\ee
The confidence diagram corresponding to this confidence matrix is shown in Figure 1. It has no closed loops and so the ${\rm Level-Slope_{2-10}}$, ${\rm Level-Mortgage}$, ${\rm Level-Corporate}$ and ${\rm Slope_{2-10}-Slope_{10-30}}$ elements of the forecasted correlation matrix can always be chosen with high confidence.

\begin{figure}
\PSbox{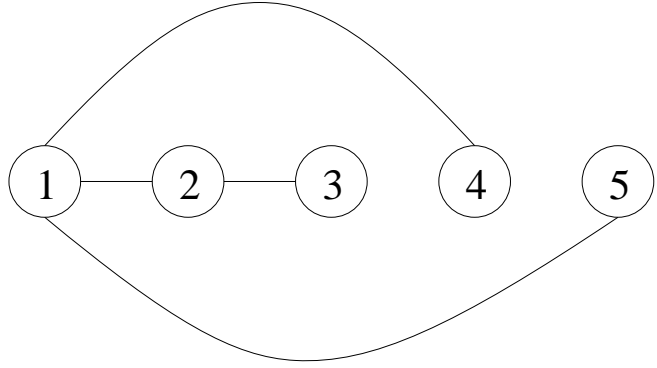 hoffset=100 voffset=10 hscale=75
vscale=75}{5.0in}{2in}
\caption{Diagram for confidence matrix in equation (\ref{weightmat}).}
\label{mA}
\end{figure}

%\newpage
% \vskip0.50in
%\begin{figure}[h]
%\begin{picture}(200,10)(-140,-20)
%\thicklines
%\put(10,0){\line(1,0){120}}
%\put(10,7){\makebox(0,0){$4$}}
%\put(50,7){\makebox(0,0){$1$}}
%\put(90,7){\makebox(0,0){$2$}}
%\put(130,7){\makebox(0,0){$3$}}
%\put(10,0){\circle*{3}}
%\put(50,0){\circle*{3}}
%\put(90,0){\circle*{3}}
%\put(130,0){\circle*{3}}
%\put(50,0){\line(0,-1){40}}
%\put(50,-40){\circle*{3}}
%\put(50,-47){\makebox(0,0){$5$}}
%\end{picture}
%\vskip0.5in
%\label{fig:prop}
%\caption{Diagram for confidence matrix in equation (\ref{weightmat}).}
%\end{figure}
\vskip0.25in
Using the confidence matrix in equation (\ref{weightmat}) the consistent correlation matrix that results from the forecasted correlation matrix appropriate to a normal environment given in equation (\ref{normalmat}) is
\be
\label{normalw}
\bordermatrix{&{\rm Level}&{\rm Slope_{2-10}}&{\rm Slope_{10-30}}&{\rm Mortgage}&{\rm Corporate}\cr
{\rm Level} & 1 & -0.500 & -0.279 & -0.250 & -0.700 \cr 
{\rm Slope_{2-10}} &-0.500 & 1 & 0.899 & 0.344 & 0.615 \cr 
{\rm Slope_{10-30}} &-0.279 & 0.899 & 1 & 0.213 & 0.273 \cr 
{\rm Mortgage} &-0.250 & 0.344 & 0.213 & 1 & 0.721 \cr 
{\rm Corporate} &-0.700 & 0.615 & 0.273 & 0.721 & 1 \cr  }.
\ee
The high confidence elements of the correlation matrix change
from their forecasted values by at most one in the third significant digit. The other elements typically change by a few in the second significant digit. Using instead
a confidence matrix where all the off diagonal elements are unity, so that none of the elements of the
forecasted correlation matrix are viewed as being more likely to be correct than any other, the consistent correlation matrix is 
\be
\label{normaluw}
\bordermatrix{&{\rm Level}&{\rm Slope_{2-10}}&{\rm Slope_{10-30}}&{\rm Mortgage}&{\rm Corporate}\cr
{\rm Level} &1 & -0.516 & -0.288 & -0.258 & -0.685 \cr 
{\rm Slope_{2-10}} &-0.516 & 1 & 0.850 & 0.331 & 0.641 \cr 
{\rm Slope_{10-30}} &-0.288 & 0.850 & 1 & 0.226 & 0.248 \cr 
{\rm Mortgage} &-0.258 & 0.331 & 0.226 & 1 & 0.722 \cr 
{\rm Corporate} &-0.685 & 0.641 & 0.248 & 0.722 & 1 \cr  }.
\ee
Now most of the elements change by a few in the second significant digit. 

Using the confidence matrix in equation (\ref{weightmat}) the consistent correlation matrix that results from the forecasted matrix appropriate to a stressed environment given in equation (\ref{stressmatnpd}) is
\be
\label{stressw}
\bordermatrix{&{\rm Level}&{\rm Slope_{2-10}}&{\rm Slope_{10-30}}&{\rm Mortgage}&{\rm Corporate}\cr
{\rm Level} &1 & -0.894 & -0.752 & -0.994 & -0.994 \cr 
{\rm Slope_{2-10}}  &-0.894 & 1 & 0.967 & 0.842 & 0.842 \cr
{\rm Slope_{10-30}} & -0.752 & 0.967 & 1 & 0.678 & 0.677 \cr 
{\rm Mortgage} &-0.994 & 0.842 & 0.678 & 1 & 1.000 \cr 
{\rm Corporate} &-0.700 & 0.842 & 0.677 & 1.000 & 1 \cr  },
\ee
and the consistent correlation matrix corresponding to a confidence matrix with all
the off diagonal elements equal to unity is
\be
\label{stressuw}
\bordermatrix{&{\rm Level}&{\rm Slope_{2-10}}&{\rm Slope_{10-30}}&{\rm Mortgage}&{\rm Corporate}\cr
{\rm Level} &1 & -0.873 & -0.565 & -1.000 & -1.000 \cr 
{\rm Slope_{2-10}} &-0.873 & 1 & 0.896 & 0.874 & 0.874 \cr 
{\rm Slope_{10-30}} &-0.565 & 0.896 & 1 & 0.566 & 0.566 \cr
{\rm Mortgage} &-1.00 & 0.874 & 0.566 & 1 & 1.000 \cr
{\rm Corporate} &-1.000 & 0.874 & 0.566 & 1.000 & 1 \cr  }.
\ee

Note that in the case where the ${\rm Slope_{2-10}-Slope_{10-30}}$ element of the
confidence matrix is 100 this element of the consistent correlation matrix
differs from its forecasted value by $-0.013=0.967-0.980$ while in the case where the ${\rm Slope_{2-10}-Slope_{10-30}}$ element of the confidence matrix is 1 this element of the consistent
correlation matrix differs from its forecasted value by $-0.084=0.896-0.980$. 
Having the ${\rm Level-Slope_{2-10}}$ and ${\rm Slope_{2-10}-Slope_{10-30}}$ elements of the confidence matrix much greater than the ${\rm Level-Slope_{10-30}}$ element leads to a large
change in the  ${\rm Level-Slope_{10-30}}$ element of the consistent correlation
matrix from its forecasted value. 

%Not every element of the
%good correlation matrix is closer to its corresponding element in the forecasted
%matrix when the confidence in its forecasted value is larger. In equation (\ref{stressuw}) the ${\rm Level-%Mortgage}$ element is equal to its forecasted value while in equation (\ref{stressw}) it differs from its %forecasted value by 0.006. However, 0.006 is still smaller than the amount 
%a typical element differs by.

The confidence matrix can contain a range of values, not just two extremes. For example
the confidence matrix
\be
\label{cfmat}
\bordermatrix{&{\rm Level}&{\rm Slope_{2-10}}&{\rm Slope_{10-30}}&{\rm Mortgage}&{\rm Corporate}\cr
	{\rm Level}			&	0	&	100	&1		&	10	&100\cr
	{\rm Slope_{2-10}}	&100		&	0	&10		&1		&1\cr
	{\rm Slope_{10-30}}	&1		&10		&	0	&1		&1 \cr
	{\rm Mortgage}		&10		&1		&1		&	0	&1 \cr
	{\rm Corporate}		&100		&1		&1		&1	& 0 },
\ee
gives
\be
\label{normalmw}
\bordermatrix{&{\rm Level}&{\rm Slope_{2-10}}&{\rm Slope_{10-30}}&{\rm Mortgage}&{\rm Corporate}\cr
{\rm Level}  &1 & -0.500 & -0.288 & -0.251 & -0.700 \cr 
{\rm Slope_{2-10}}	&-0.500 & 1 & 0.893 & 0.343 & 0.619\cr 
{\rm Slope_{10-30}} &-0.288 & 0.893 & 1 & 0.210 & 0.263 \cr
{\rm Mortgage} & -0.251 & 0.343 & 0.210 & 1 & 0.715 \cr
{\rm Corporate} &-0.700 & 0.619 & 0.263 & 0.715 & 1 \cr },
\ee
for the consistent correlation matrix corresponding to the forecasted correlation
matrix in equation (\ref{normalmat}) appropriate to normal environments and
\be
\label{stressmmw}
\bordermatrix{&{\rm Level}&{\rm Slope_{2-10}}&{\rm Slope_{10-30}}&{\rm Mortgage}& {\rm Corporate}\cr
{\rm Level}  &1 & -0.897 & -0.673 & -0.992 & -0.997 \cr 
{\rm Slope_{2-10}}	&-0.897 & 1 & 0.931 & 0.851 & 0.860\cr 
{\rm Slope_{10-30}} &-0.673 & 0.931 & 1 & 0.601 & 0.613 \cr
{\rm Mortgage} & -0.992 & 0.851 & 0.601 & 1 & 0.998 \cr 
{\rm Corporate} &-0.997 & 0.860 & 0.613 & 0.998 & 1 \cr},
\ee
for the consistent correlation matrix corresponding to the stressed environment forecast in equation (\ref{stressmatnpd}).

 For the stressed scenario, the inconsistency in the forecasted matrix was easy to understand. The  $100$ elements of the confidence matrix in equation (\ref{weightmat}) force the ${\rm Level-Slope_{2-10}}$ and ${\rm Slope_{2-10}-Slope_{10-30}}$ elements to be near their forecasted values. In this case the only way the forecasted matrix can change to
a consistent one is via a large shift in its ${\rm Level-Slope_{10-30}}$ element, from $-0.55$ to $-0.752$. When the confidence matrix contains only $1$'s on the off diagonal such an extreme change in one of its elements is not favored and the change in the ${\rm Level-Slope_{10-30}}$ element is only from $-0.55$ to $-0.565$. In this case a larger change occurs in the ${\rm Slope_{2-10}-Slope_{10-30}}$ element which goes from $0.98$ to $0.896$. Finally for the confidence matrix in equation (\ref{cfmat}) the results are between these two extremes. The  ${\rm Level-Slope_{10-30}}$ changes from $-0.55$ to $-0.673$ and the ${\rm Slope_{2-10}-Slope_{10-30}}$ changes from $0.98$ to $0.931$.

Our parameterization which for $N \times N$ correlation matrices involves only $N (N-1)/2$ angles should
yield a faster numerical minimization than the parameterization of Rebonato and J\"{a}ckel which uses $N (N-1)$ angles. The numerical minimization of the function $f$ can be sped up by a judicious choice of starting values for the angles that specify the consistent correlation matrix. For example,
taking the initial values of the angles as those that correspond to the consistent correlation matrix formed using the spectral decomposition method outlined in section 3 of Rebonato and J\"{a}ckel (2000).

\section{From Correlations to Risk}

The market price of any portfolio is assumed to be completely dependent on the factors that are the sources of risk. 
For a portfolio $P(L,S_{2-10},S_{10-30},M,C)$ the absolute sensitivity to the factors is
\be
dP= {\partial P \over \partial L} dL + 
{\partial P \over \partial S_{2-10}} dS_{2-10} +
{\partial P \over \partial S_{10-30}} dS_{10-30} +
{\partial P \over \partial M} dM +
{\partial P \over \partial C} dC .
\ee
The sensitivity of the portfolio with respect to a given factor can be deduced from this. The duration of the portfolio with respect to level is simply

\be
-{1 \over P} {dP \over dL} = -{1 \over P} \left [ {\partial P \over \partial L} + {\partial P \over \partial S_{2-10}} {\partial S_{2-10} \over \partial L}  + {\partial P \over \partial S_{10-30}} {\partial S_{10-30} \over \partial L} + {\partial P \over \partial M} {\partial M \over \partial L} + {\partial P \over \partial C} {\partial C \over \partial L} \right ],
\ee
which can be written in terms of ``betas". Suppose the change in a spread factor $S$ depends on the change in the factor $L$ by the simple linear regression, 
\be
dS=dX_S+\beta_{dS,dL}dL.
\ee
Here $dX_S$ is not correlated with $dL$. Multiplying both sides of this equation by $dL$, and averaging we get, $E[dS \cdot dL]  = \beta_{dS,dL}E[dL \cdot dL]$. Since $E[dS \cdot dL] = \rho_{dS,dL} \sigma_{dS} \sigma_{dL}$ and $E[dL \cdot dL] = \sigma_{dL}^2$
we obtain,
\be
 \beta_{dS,dL} = \rho_{dS,dL} {\sigma_{dS} \over \sigma_{dL}}.
\ee
Repeating the derivation for all the other betas in terms of correlations and volatilities we obtain that the total duration of a portfolio (with respect to level) is
\bea
-{1 \over P} {dP \over dL} &=&-{1 \over P} \left[ {\partial P \over \partial L} 
+\rho_{dS_{2-10},dL} {\sigma_{dS_{2-10}} \over \sigma_{dL}} {\partial P \over \partial S_{2-10}} \right. \nonumber \\ 
&+& \left. \rho_{dS_{10-30},dL} {\sigma_{dS_{10-30}} \over \sigma_{dL}} {\partial P \over \partial S_{10-30}} 
+\rho_{dM,dL} {\sigma_{dM} \over \sigma_{dL}} {\partial P \over \partial M} 
+\rho_{dC,dL} {\sigma_{dC} \over \sigma_{dL}} {\partial P \over \partial C} 
\right].
\eea

Relabeling the partial (and total) durations with the letter $D$, and comparing
with a benchmark portfolio we can write the over or under exposure of the total portfolio
duration (with respect to level) as,
\bea
\label{keyrisk}
\Delta D^{(L)}_{Total}&=& \Delta D_L + \rho_{dS_{2-10},dL} {\sigma_{dS_{2-10}} \over \sigma_{dL}} \Delta D_{2-10} + \rho_{dS_{10-30},dL} {\sigma_{dS_{10-30}} \over \sigma_{dL}} \Delta D_{10-30}\nonumber\\ 
&+&\rho_{dM,dL} {\sigma_{dM} \over \sigma_{dL}}\Delta D_M +\rho_{dC,dL} {\sigma_{dC} \over \sigma_{dL}} \Delta D_C, 
\eea
where $\Delta$ signifies the difference in partial or total duration of the portfolio's exposure versus the index exposure.
The last equation is the key result for computing effective duration risk of portfolios due to correlation of risk factors. One could use historical estimates of correlations for the $\rho$'s, or pick them from consistent forecast correlation matrices, as we propose in this paper.

Reasonable measures for the annualized volatilities (in basis points) are: $\sigma_{dL} = 100$, $\sigma_{dS_{2-10}} =75$, $\sigma_{dS_{10-30}} = 35$, $\sigma_{dM} =25$ and $\sigma_{dC} = 50$ and we use these values for the remainder of this section. Assume that the overweights and underweights of a representative fixed income portfolio versus the index are: $\Delta D_L = 1.00$, $\Delta D_{2-10}=0.30$, $\Delta D_{10-30} = 0.50$, $\Delta D_M =1.50$ and $\Delta D_C =-0.50$. Using these values   
the contributions of the ${\rm Level}$, ${\rm Slope_{2-10}}$, ${\rm Slope_{10-30}}$, ${\rm Mortgage}$ and ${\rm Corporate}$ partial durations to the total duration (with respect to level), adjusted for the correlation matrix in equation (\ref{normalmw}) respectively are $1.00, -0.11, -0.05, -0.09$ and $0.17$  giving a total duration of $0.92$. Doing the same with the stressed correlation matrix in equation (\ref{stressmmw}) gives the partial contributions $1.00, -0.20, -0.12, -0.37$ and $0.25$ whose sum is only $0.56$.  The forecasted stress scenario correlation matrix in equation (\ref{stressmatnpd}) corresponds to a portfolio that is almost half a year shorter! The practical import of this for portfolio management is clear - if we believe that in periods of stress the correlations between some of the market factors would go to the extreme levels as proposed, we need to be extra long by almost half a year to insure having enough long exposure during the stress scenario. For portfolios that have high negative convexity (for example due to mortgages), this shortening bias in stressed scenarios is on top of the shortening due to negative convexity, and solely due to correlations.

A useful measure of stress risk is the change in portfolio total level duration as we go from the normal environment to the stressed environment.  This is a very simple way of identifying portfolios that are likely to become performance outliers in an environment of stress. Furthermore equation (\ref{keyrisk}) for the total level duration is independent of the statistics of the risk factors\footnote{It does however depend on the
linear regression.} ({\it i.e.} they don't have to be normal). To illustrate this approach, we create forty eight illustrative portfolios by varying the five key risk measures. The selection of the different combinations was done on the following basis: (a) three types of portfolio duration with respect to level - extreme overweight (1 year), moderate overweight (half a year), and flat (no duration overweight); (b) two types of portfolios based on $2-10$ duration, one with a steepening bias and the other with no bias; (c) two types of portfolios based on
$10-30$ duration one with no bias and the other with a steepening bias; (d) two types of portfolios based on mortgage spread duration - high ($1.5$ year overweight) and flat (no overweight); (e) two types of portfolios based on corporate spread duration - neutral (no underweight) and underweight ($-0.50$ year underweight). This gives us $48$ portfolio combinations. Using the correlation matrices in equation (\ref{normalmw}) and equation (\ref{stressmmw}) the total level durations of these portfolios are shown in Tables I and II.

The last three columns of these tables respectively give the total level duration in the normal environment, stressed environment and its change. The correlation matrices in equation (\ref{normalmw}) and equation (\ref{stressmmw}) were used and they usually make the total duration lower than the partial level duration in both the normal and stressed cases. However, what is striking is that the durations can get shortened by almost fifty percent in going from normal to stressed scenarios if there is a large mortgage spread duration exposure. For example, portfolio number 22 has almost zero total duration in the 
normal environment, but because of high mortgage spread duration exposure, it can actually become short the market by more than a quarter year in the stressed environment.

The consistent correlation matrices obtained by the procedure we have described can be used directly in computing the total risk of the portfolio as well \footnote{We appreciate the editor's comment regarding this.}. The variance of the portfolio is  $\sum \Delta D_i \sigma_i \rho_{ij} \sigma_j \Delta D_j$, where the indices $i$ and $j$ label the risk factors and are summed over. There are two reasons we have chosen to illustrate the impact of correlations on total level duration instead of total variance in this paper - both of which result from our view on standard industry practice for managing fixed income portfolios. First, most fixed income portfolios are immunized separately with respect to their sensitivities to different risk factors, and condensing risk information into one number such as total variance is likely to adversely impact the multi-dimensional view of risk. Second, very generally speaking, total level duration is usually the primary risk exposure for large fixed income portfolios, since level shifts of the yield curve can impact the portfolio's performance much more rapidly and irreversibly than the secondary factors such as curve, spread etc. 

\newpage
\vskip 0.5in
\noindent
Table I. Sample Portfolios (with $\Delta D_{10-30}$=0): Change in Total Level Duration from Normal to Stressed Environments using the correlation matrices in equations (\ref{normalmw}) and (\ref{stressmmw}).
\vskip 0.25in
\noindent
$
\bordermatrix{&\Delta D_L&\Delta D_{2-10}&\Delta D_{10-30}&\Delta D_M&\Delta D_C &Normal~ \Delta D^{(L)}_{Total} & Stressed~ \Delta D^{(L)}_{Total} & Change \cr
P1 & 1 & 0.3 & 0 & 1.5 & -0.5 & 0.968 & 0.675 & -0.293 \cr 
P2 & 1 & 0.3 & 0 & 1.5 & 0 & 0.793 & 0.426 & -0.367 \cr 
P3 &1 & 0.3 & 0 & 0 & -0.5 & 1.063 & 1.047 & -0.015 \cr
P4 & 1 & 0.3 & 0 & 0 & 0 & 0.888 & 0.799 & -0.089 \cr
P5 & 1 & 0 & 0 & 1.5 & -0.5 & 1.081 & 0.877 &    -0.204 \cr
P6 & 1 & 0 & 0 & 1.5 & 0 & 0.906 & 0.628 & -0.278 \cr
P7 & 1 & 0 & 0 & 0 & -0.5 & 1.175 & 1.249 &  0.074 \cr
P8 & 1 & 0 & 0 & 0 & 0 & 1 & 1 & 0 \cr 
P9 &0.5 & 0.3 & 0 & 1.5 & -0.5 & 0.468 & 0.175 & -0.293 \cr 
P10&0.5 & 0.3 & 0 & 1.5 & 0 & 0.293 &  -0.074 & -0.367 \cr
P11& 0.5 & 0.3 & 0 & 0 & -0.5 & 0.563 & 0.547 & -0.015 \cr 
P12&0.5 & 0.3 & 0 & 0 & 0 & 0.388 & 0.298 & -0.089 \cr 
P13&0.5 & 0 & 0 & 1.5 & -0.5 & 0.581 & 0.377 & -0.204 \cr 
P14&0.5 & 0 & 0 & 1.5 & 0 & 0.406 & 0.128 & -0.278 \cr
P15& 0.5 & 0 & 0 & 0 & -0.5 & 0.675 & 0.749 & 0.074 \cr 
P16&0.5 & 0 & 0 & 0 & 0 & 0.5 & 0.5 & 0. \cr
P17& 0 & 0.3 & 0 & 1.5 & -0.5 & -0.032 & -0.325 & -0.293 \cr 
P18&0 & 0.3 & 0 & 1.5 & 0 & -0.207 & -0.574 & -0.367 \cr 
P19&0 & 0.3 & 0 & 0 & -0.5 & 0.063 & 0.047 & -0.015 \cr 
P20&0 & 0.3 & 0 & 0 & 0 & -0.113 & -0.202 & -0.089 \cr
P21& 0 & 0 & 0 & 1.5 & -0.5 & 0.081 & -0.123 & -0.204 \cr 
P22&0 & 0 & 0 & 1.5 & 0 & -0.094 & -0.372 & -0.278 \cr 
P23&0 & 0 & 0 & 0 & -0.5 & 0.175 & 0.249 & 0.074 \cr 
P24&0 & 0 & 0 & 0 & 0 & 0 & 0 & 0 \cr  }
$

\newpage
\noindent
Table II. Sample Portfolios (with $\Delta D_{10-30}$=0.5):  Change in Total Level Duration from Normal to Stressed Environments using the correlation matrices in equations (\ref{normalmw}) and (\ref{stressmmw}).
\vskip 0.25in
\noindent
$
\bordermatrix{&\Delta D_L&\Delta D_{2-10}&\Delta D_{10-30}&\Delta D_M&\Delta D_C &Normal~ \Delta D^{(L)}_{Total} & Stressed~ \Delta D^{(L)}_{Total} & Change \cr
P25 & 1 & 0.3 & 0.5 & 1.5 & -0.5 & 0.918 & 0.558 & -0.360 \cr 
P26 &1 & 0.3 & 0.5 & 1.5 & 0 & 0.743 & 0.308 & -0.435 \cr 
P27 &1 & 0.3 & 0.5 & 0 & -0.5 & 1.012 & 0.930 & -0.082 \cr
P28 & 1 & 0.3 & 0.5 & 0 & 0 & 0.837 & 0.680 & -0.157\cr 
P29 & 1 & 0 & 0.5 & 1.5 & -0.5 & 1.030 & 0.759 &  -0.271 \cr 
P30 & 1 & 0 & 0.5 & 1.5 & 0 & 0.855 & 0.510 & -0.345 \cr 
P31 & 1 & 0 & 0.5 & 0 & -0.5 & 1.125 & 1.131 & 0.007 \cr 
P32 & 1 & 0 & 0.5 & 0 & 0 &  0.950 & 0.882 & -0.067 \cr 
P33 & 0.5 & 0.3 & 0.5 & 1.5 & -0.5 & 0.418 & 0.058 & -0.360 \cr
P34 & 0.5 & 0.3 & 0.5 & 1.5 & 0 & 0.243 & -0.192 & -0.435 \cr
P35 & 0.5 & 0.3 & 0.5 & 0 & -0.5 & 0.512 & 0.430 & -0.082 \cr 
P36 & 0.5 & 0.3 & 0.5 & 0 & 0 & 0.337 & 0.180 & -0.157 \cr 
P37 & 0.5 & 0 & 0.5 & 1.5 & -0.5 & 0.530 & 0.259 & -0.271 \cr 
P38 &0.5 & 0 & 0.5 & 1.5 & 0 & 0.355 & 0.010 & -0.345 \cr 
P39& 0.5 & 0 & 0.5 & 0 & -0.5 & 0.625 & 0.631 & 0.007 \cr 
P40&0.5 & 0 & 0.5 & 0 & 0 & 0.450 & 0.382 & -0.067 \cr 
P41&0 & 0.3 & 0.5 & 1.5 & -0.5 & -0.082 & -0.442 &  -0.360 \cr 
P42&0 & 0.3 & 0.5 & 1.5 & 0 & -0.257 & -0.692 & -0.435 \cr
P43& 0 & 0.3 & 0.5 & 0 & -0.5 & 0.012 & -0.070 &  -0.082 \cr 
P44&0 & 0.3 & 0.5 & 0 & 0 & -0.163 & -0.320 & -0.157 \cr
P45& 0 & 0 & 0.5 & 1.5 & -0.5 & 0.030 & -0.241 & -0.271 \cr 
P46&0 & 0 & 0.5 & 1.5 & 0 & -0.145 & -0.490 & -0.345 \cr
P47& 0 & 0 & 0.5 & 0 & -0.5 & 0.125 & 0.131 & 0.007 \cr
P48& 0 & 0 & 0.5 & 0 & 0 & -0.050 & -0.118 & -0.067 \cr  }
$

\section{Concluding Remarks}

The use of historical correlations for predicting portfolio risk is a dubious procedure. In times of market stress such estimates can be extremely misleading and even in normal market environments can be quite inaccurate. An alternate procedure is to focus on a few of the key risk factors and forecast their correlations (and volatilities) based on current and expected future economic conditions. However, a correlation matrix constructed in this way will not always be mathematically consistent. If it isn't positive semi-definite then it allows negative variance, which is impossible. 

In general, a group of  portfolio managers are likely to have a high degree of confidence in forecasting specific elements of the correlation matrix, e.g. the correlation between level and slope, and slightly lower degree of confidence in forecasting other correlations, {\it e.g.} the correlation between level and spreads. The higher degree of
confidence could arise from a strong macroeconomic view, or from a consensus among a number of portfolio managers. 
We have elaborated on the methodology proposed recently by Rebonato and J\"{a}ckel (2000) that uses this ranking to obtain a mathematically consistent correlation matrix that is closest to the original forecasted one. One improvement on their work in the use of rotational invariance
to specify the most general mathematically consistent correlation matrix in terms of only $N (N-1)/2$ angles.

In times of market stress correlations of risk factors have a tendency to quickly evolve to values that are well beyond their historical values. To emphasize the importance of forecasting correlations to asses the risk in such environments we showed in some simple examples that the increase in correlations can drastically change the total level duration of a portfolio. 

\section{Acknowledgments}
We would like to thank colleagues at PIMCO for their critical reading and numerous detailed comments on an initial version of this manuscript. One of us (VB) would like to thank Chris Dialynas for enlightening discussions on past unpublished work on related topics.

\vspace{0.4cm}

\noindent
{\Large{\bf References}}

\vspace{0.4cm}

\noindent
V. Bhansali (1997), {\it Pricing and Managing Exotic and Hybrid Options}, McGraw-Hill.

\vspace{0.3cm}

\noindent
B.H. Boyer, M.S. Gibson and M. Loretan (1997), {\it Pitfalls in tests for changes in correlations}, International Finance
Discussion Paper No. 597, December.

\vspace{0.3cm}

\noindent
J.M. Campa and K. Chang (1998), {\it The Forecasting Ability of Correlations Implied in Foreign Exchange Options}, Journal of International Money and Finance, 17.

\vspace{0.3cm}

\noindent
C. Finger (1997), {\it A methodology for stress correlation}, In: Risk Metrics Monitor, Fourth Quarter.

\vspace{0.3cm}

\noindent
P. Jorion (2001), {\it Value at Risk}, Second Edition, McGraw-Hill.

\vspace{0.3cm}

\noindent
P.H. Kupiec (1998), {\it Stress testing in a value at risk framework}, Journal of Derivatives, Volume 6.

\vspace{0.3cm}

\noindent
R. Litterman and K. Winkelmann (1998), {\it Estimating Covariance Matrices}, Goldman Sachs.

\vspace{0.3cm}

\noindent
R. Rebonato and P. J\"{a}ckel (2000), {\it The most general methodology for creating a valid correlation matrix for risk management and option pricing purposes}, The Journal of Risk, Volume 2, Number 2.

\vspace{0.3cm}

\noindent
P. Wilmott (2000), {\it Quantitative Finance}, John Wiley and Sons.

\vspace{0.3cm}

\noindent
J. Xiongwei and N. Pearson (1999), {\it  Using Value at Risk to Control Risk Taking: How wrong can you be?}, The Journal of
Risk, Volume 1, Number 2.

\end{document}